\documentclass[a4]{article}

\usepackage[utf8]{inputenc}
\usepackage[T1]{fontenc}
\usepackage[english]{babel}
\usepackage{amsmath,amssymb,amsthm}
\usepackage{tikz-cd}
\usepackage{graphicx} % Required for inserting images
\usepackage[margin=1.in]{geometry}
\usepackage[shortlabels]{enumitem}
\usepackage{todonotes}
\usepackage[bibstyle=numeric, backend=biber, sorting=none, citestyle=numeric-comp, giveninits,url=false,maxbibnames=5]{biblatex} % loads etoolbox
\usepackage{csquotes}\MakeOuterQuote"
\usepackage{soul}\setstcolor{red}
\usepackage[colorlinks,citecolor=blue,linkcolor=blue]{hyperref}
\usepackage[capitalize]{cleveref}
\usepackage{comment}
\usepackage{stackengine}

\DeclareUnicodeCharacter{0141}{\Lbar{}}

\newtheorem{theorem}{Theorem}
\newtheorem{definition}[theorem]{Definition}

\theoremstyle{definition}

\newtheorem*{problem*}{Problem}
\newtheorem*{assumption*}{Assumption}

\newtheorem*{warning*}{Warning}

\newcommand{\ip}[2]{\langle #1,#2\rangle}

\newcommand{\norm}[1]{\lVert #1\rVert}

\newcommand{\ox}{\otimes}
\newcommand{\mc}{\mathcal}
\newcommand{\eps}{\varepsilon}

\newcommand{\II}{{\mathrm{II}}}

\newcommand{\abs}[1]{\lvert #1 \rvert}

\DeclareMathOperator{\tr}{tr}
\newcommand{\hide}[1]{}

\def\A{{\mc A}}

\def\CC{{\mathbb C}}

\def\H{{\mc H}}

\def\K{{\mathcal K}}

\def\U{{\mc U}}

\newcommand{\Lbar}{\L{}}

\usepackage{orcidlink}

\title{Uniqueness of purifications is equivalent to Haag duality}
\author{Lauritz van Luijk\,\orcidlink{0000-0003-3153-549X}, Alexander Stottmeister\,\orcidlink{0000-0002-0145-0877}, Henrik Wilming\,\orcidlink{0000-0002-0306-7679}}
\date{\small Leibniz Universit\"at Hannover, Institut f\"ur Theoretische Physik, Appelstraße 2, 30167 Hannover, Germany\\  \today}

\begin{document}

\maketitle
\begin{abstract}
The uniqueness of purifications of quantum states on a system $A$ up to local unitary transformations on a purifying system $B$ is central to quantum information theory.  We show that, if the two systems are modelled by commuting von Neumann algebras $M_A$ and $M_B$ on a Hilbert space $\H$, then uniqueness of purifications is equivalent to \emph{Haag duality} $M_A = M_B'$. In particular, the uniqueness of purifications can fail in systems with infinitely many degrees of freedom---even when $M_A$ and $M_B$ are commuting factors that jointly generate $B(\H)$ and hence allow for local tomography of all density matrices on $\H$. 
\end{abstract}

\section{Introduction}

Recently, motivated by many-body physics, quantum field theory, and quantum gravity, rigorous results on entanglement in systems with infinitely many degrees of freedom have been receiving increased interest \cite{matsui_split_2001,keyl_infinitely_2003,keyl_entanglement_2006,verch_distillability_2005,matsui_boundedness_2013,hollands_entanglement_2018,wittenAPSMedalExceptional2018,cleve_constant_2022,crann_state_2020,van_luijk_schmidt_2024,van_luijk_pure_2024,van_luijk_critical_2025}.
When describing entanglement in such systems, an operator algebraic approach offers a natural, model-free perspective.
In this setting, there is, in general, no tensor product decomposition of the Hilbert space describing subsystems.
Instead, subsystems $A,B,\ldots$ correspond to von Neumann algebras $M_A,M_B,\ldots$ on the full system's Hilbert space $\H$.
Importantly, the von Neumann algebras $M_A$ and $M_B$ commute if the corresponding subsystems are distinct.

If both subsystems are purely quantum, i.e., do not contain degrees of freedom that can be measured without being perturbed, then $M_A$ and $M_B$ are \emph{factors}, see \cite{kuramochi_accessible_2018} or \cite[Sec.~4.4]{vanluijkEntanglementNeumannAlgebraic2025}.
A factor is a von Neumann algebra $M\subset B(\H)$ such that $M \cap M' = \CC$, where $M'$ denotes the commutant: $M' = \{x\in B(\H) : [x,y]=0\ \forall y\in M\}$.
    
A typical situation that one considers is that of a bipartition: The full system is divided into the two parts $A$ and $B$ so that no degrees of freedom are left out.
This immediately leads to our central question:
\begin{center}\textit{
    How does one model the fact that no degree of freedom is left out?
    }
\end{center}

Below we will discuss three different ways to model this requirement: 1) Local tomography, 2) Haag duality, and 3) the \emph{Uhlmann property}, encoding the uniqueness of purifications. In finite dimensions, more generally in a tensor product framework, all of these are equivalent. 
In systems with infinitely many degrees of freedom, this equivalence generally fails, motivating to clarify the relation between different models. 
For purely quantum systems, Haag duality implies local tomography, but the converse is false.
Our main result is that Haag duality and the Uhlmann property are equivalent. 

We discuss below how these different ways of modeling bipartite systems play out in spatial bipartitions of quantum many-body systems \emph{described by pure states}.
The upshot of this discussion is that local tomography is automatic, whereas Haag duality is not.
Instead, the validity of Haag duality depends on the physics of the systems and on the geometry of the bipartition.
Verifying it in concrete models is a notoriously difficult task \cite{matsui_split_2001,keyl_entanglement_2006,naaijkens_haag_2012,matsui_boundedness_2013,fiedler_haag_2015, van_luijk_critical_2025}.

\subsection{Local Tomography}
A (seemingly) minimal requirement that $A$ and $B$ together constitute the full system is that every density operator $\rho$ on $\H$ should allow for \emph{local tomography} by Alice and Bob: It can be uniquely identified by all correlation functions of the form 
\begin{align}
    (a,b) \mapsto \tr(\rho ab),  \qquad a\in M_A,\ b\in M_B.
\end{align}
The local tomography condition is equivalent to $M_A \vee M_B = B(\H)$, where $M_A\vee M_B = (M_A\cup M_B)''$ is the von Neumann algebra generated by $M_A$ and $M_B$. Since $M_A$ and $M_B$ commute, local tomography requires that $M_A$ and $M_B$ are factors:
\begin{align}\label{eq:factor}
    M_A \cap M_A' \subseteq M_B' \cap M_A' = (M_B \vee M_A)' = B(\H)' = \CC,
\end{align}
and similarly $M_B \cap M_B'=\CC$.
Indeed, if $M_A$ is not a factor, there is a non-trivial unitary operator $u$ that commutes with both $M_A$ and $M_B$. Hence, there are vectors $\Psi\neq u\Psi$ which cannot be distinguished by local tomography.
To summarize:
Local tomography holds if and only if $M_A$ and $M_B$ are factors such that $M_A\vee M_B=B(\H)$ (or, equivalently, $M_A'\cap M_B'=\CC$).

\subsection{Haag duality}
Instead of the local tomography condition, one could also consider a different way to model that no degrees of freedom are left out: 
Demanding that Bob has access to all the operators that commute with all of Alice's operators. Mathematically, this means that 
\begin{align}
    M_B = M_A'.
\end{align}
This condition is known as \emph{Haag duality}, after a related property in algebraic quantum field theory \cite{haag_local_1996,keyl_infinitely_2003}.

Haag duality (or approximate versions of it) is of vital importance to derive rigorous results on entanglement in systems with infinitely many degrees of freedom \cite{summers_vacuum_1985,summers_maximal_1988,crann_state_2020,van_luijk_pure_2024,van_luijk_embezzlement_2024,keyl_infinitely_2003,verch_distillability_2005} and in the rigorous treatment of topologically ordered systems in the thermodynamic limit \cite{naaijkensLocalizedEndomorphismsKitaevs2011,naaijkens_haag_2012, naaijkens_kosaki-longo_2013, naaijkens_subfactors_2018, jones_dhr_2024, jones_local_2023, fiedler_jones_2017, fiedler_haag_2015, chuah_boundary_2024,bhardwaj_superselection_2025}. 
Often considered a technical condition \cite{keyl_entanglement_2006,fiedler_jones_2017,naaijkensQuantumSpinSystems2017,bols_category_2025,bhardwaj_superselection_2025}, it has lacked a clear operational interpretation in terms of quantum information theory so far. 

If $M_A$ is a factor and Haag duality holds, then $M_B$ is a factor, too.
In this case, Haag duality implies local tomography (and only in this case, since local tomography requires factors). 
But the converse is false in general: There exist pairs of factors $M_A,M_B$ that allow for local tomography, but the inclusion $M_A \subset M_B'$ is strict. Notably, there is no non-trivial operator in $M_B'$ that commutes with $M_A$:
Since $M_A\vee M_B=B(\H)$ by local tomography we have (see \cref{eq:factor})
\begin{align}
    M_A' \cap M_B' = \CC.
\end{align}
Such a situation is termed an \emph{irreducible subfactor inclusion} in mathematics and is a rich topic of research \cite{jones_index_1983,jones_introduction_1997,longo_index_1989,longo_index_1990,longo_theory_1997,longo_notes_2001,kosaki1998type,kosaki_extension_1986,bisch_new_2025}. We will discuss a physical example below. 
We conclude that Haag duality cannot be given an operational interpretation in terms of local tomography. Of course, Haag duality is trivially true in the tensor product framework, where $\H=\H_A\ox \H_B$, $M_A = B(\H_A)\ox 1$ and $M_B = 1\ox B(\H_B)$.

\subsection{Uhlmann property} 
It is of fundamental importance to quantum information theory that every density matrix has a purification on a larger Hilbert space and that all such purifications are related by a (partial) isometry between the purifying Hilbert spaces. 
Purifications are unique up to (partial) isometries.
Operationally, this means that if $\rho$ on $\H_A$ describes Alice's subsystem and $\Psi_\rho$ is a purification on $\H = \H_A\otimes \H_B$, then Bob, acting on $\H_B$ by local operations, can prepare any other purification of Alice's state.

An important consequence of the uniqueness of purifications is \emph{Uhlmann's theorem} \cite{uhlmann_transition_1976,nielsen_quantum_2010}, which characterizes the Fidelity $F(\rho,\sigma)=\norm{\rho^{1/2}\sigma^{1/2}}_1$ of a pair of density operators $\rho,\sigma$ on a Hilbert space $\H$ as the maximal overlap of their purifications
\begin{equation}\label{eq:Uhlmann}
    F(\rho,\sigma) = \sup_U\  \abs{\ip{\Psi_\sigma}{(1\ox U)\Psi_\rho}},
\end{equation}
where $\Psi_\rho$ and $\Psi_\sigma$ are two arbitrary purifications in a common extension $\H\ox \K$ and where the optimization is over all unitaries $U$ on $\K$. 
Uhlmann's theorem is central to quantum information theory. 
Recently, it has received renewed interest from the point of view of the quantum computational complexity \cite{metgerStateQIPStatePSPACE2023, bostanciUnitaryComplexityUhlmann2025, utsumiQuantumAlgorithmsUhlmann2025a} as well as from the point of view of its rigidity \cite{bostanciLocalTransformationsBipartite2025}.

As stated above, Uhlmann's theorem only applies to systems with finitely many degrees of freedom.
The last condition we consider to formalize that no degree of freedom is left out is whether uniqueness of purifications (hence Uhlmann's theorem) is fulfilled: Starting from any pure state $\Psi$ on $\H$, is it true that any other $\Phi \in \H$ with the same reduced state on $M_A$ as $\Psi$ may be reached by acting with local operations by Bob? We formalize this property as follows:

\begin{definition}[Uhlmann property]\label{def:uhlmann}
    A pair of of commuting von Neumann algebras $M_A,M_B$ on $\H$ has the \emph{Uhlmann property} if for any two pure state vectors $\Phi,\Psi\in \H$ such that $\langle \Psi, a\Psi \rangle = \langle \Phi,a\Phi\rangle$ for all $a\in M_A$ implies that for any $\eps>0$ there exists a unitary $u\in M_B$ such that $|\langle \Psi, u\Phi\rangle| \geq 1-\eps$.
\end{definition}

We emphasize that the Uhlmann property only makes a statement about those states on $M_A$ that admit a purification on the joint system, but does not require that all reduced states of one of the subsystems have purifications.

In contrast to local tomography and Haag duality, the Uhlmann property is \emph{a priori} not symmetric relative to the subsystems $A$ and $B$. It will be a consequence of our main result that it is, in fact, symmetric.

In the definition of the Uhlmann property, we already anticipate that in systems with infinitely many degrees of freedom, one can generally only approximate a given purification $\Phi$ up to arbitrary accuracy using unitary operators from $M_B$ acting on $\Psi$. 
However, we will see below that if this is possible, then one can in fact find a local partial isometry, and hence a local quantum channel, that deterministically maps $\Psi$ to $\Phi$.

The approximate nature of \cref{def:uhlmann} has the important consequence that we can freely restrict the unitaries to any weakly-dense subgroup $G_B \subset \U(M_B)$ of the unitary group $\U(M_B)$, without changing the validity of the Uhlmann property.
This is, for instance, useful when discussing quantum many-body systems, where we can take $G_B$ to the group of finitely supported unitaries in the region $B$ (see \cref{sec:many-body} below).

\subsection{Bipartitions of quantum many-body systems in the thermodynamic limit}\label{sec:many-body}
An important motivation for studying entanglement in operator algebraic settings is given by many-body physics at zero temperature and in the thermodynamic limit. We briefly review their description, see, for example, \cite{naaijkensQuantumSpinSystems2017} for details.
Consider an infinite lattice $\Gamma$ of spins with local Hilbert space dimension $d$. 
Let $\A_\Gamma$ be the *-algebra of operators that have finite support on $\Gamma$.
For any subset $A\subset \Gamma$, let $\A_A\subset \A_\Gamma$ be the *-subalgebra of operators with finite support in $A$.
If $A_1\subset A_2$, then $\A_{A_1}\subset \A_{A_2}$ and if $A_1 \cap A_2 =\emptyset$, then $\A_{A_1}$ and $\A_{A_2}$ commute.
If $\Gamma = A\cup B$ is a bipartition of the lattice, we have $\A_\Gamma \cong \A_A\odot \A_B$, where "$\odot$" denotes the algebraic tensor product.
Thus, the analog of Haag duality always holds on the level of operators with finite support: If $x\in \A_\Gamma$ commutes with $\A_A$, then $x\in \A_B$.

A ground state of a Hamiltonian is an algebraic state $\omega$ on $\A_\Gamma$. 
The ground state sector $\H$ arises from its GNS representation $\pi$, and for any bipartition $A\cup B = \Gamma$ we obtain von Neumann algebras $M_A=\pi(\A_A)''$ and $M_B=\pi(\A_B)''$.
If $\omega$ is a pure state, $\pi$ is irreducible and we automatically have that $M_A \vee M_B = B(\H)$. 
Thus, for any bipartition $\Gamma = A\cup B$, the factors $(M_A,M_B)$ satisfy the local tomography criterion.
However, in general, Haag duality $M_A= M_B'$ fails---although the relative analogue of it, $\A_{A}'\cap \A_{\Gamma} = \A_{B}$, holds on the level of finitely supported operators, as discussed above.

We will see an explicit example of such a failure in \cref{sec:example}. As mentioned in the introduction, proving Haag duality in specific situations is generally difficult and no general method is available; see \cite{matsui_split_2001,keyl_entanglement_2006,naaijkens_haag_2012,matsui_boundedness_2013,fiedler_haag_2015, van_luijk_critical_2025} for examples. 

The unitary group of $\A_B$ is weakly dense in $\U(M_B)$.\footnote{Consider an increasing sequence of finite subsystems $B_k \subset B_{k+1}$ such that $B=\cup_{k=1}^\infty B_k$. Since each $\A_{B_k}$ is a finite-dimensional matrix algebra, it is a von Neumann algebra. We have $\A_B = \cup_k \A_{B_k}$, hence $M_B = (\cup_k \A_{B_k})''$ and it follows from Kaplansky's theorem \cite[Thm.~II.4.8]{takesaki1} that $\U(M_B)$ is the strong*-closure of $\U(\A_{B}) = \cup_k \U(\A_{B_k}$, see for example \cite[Supplemental Material, Lemma 38]{van_luijk_critical_2025}. In particular, the group $\U(\A_B)$ of finitely localized unitaries are weakly dense in $\U(M_B)$.}
Hence, by the comments at the end of the previous section, Uhlmann's property can be completely formulated in terms of unitary operators with finite support. Thus, if Haag duality holds, $B$ can approximately map any purification to any other purification with unitary operators from $\A_B$.

\clearpage

\section{Main result}

We are now in the position to state our main result.
\begin{theorem}\label{thm:main} 
    A pair of commuting von Neumann algebras $M_A, M_B$ on $\H$ has the Uhlmann property if and only if Haag duality holds.
    In this case, if state vectors $\Psi,\Phi\in \H$ have the same $A$-marginal, there exists a partial isometry $v \in M_B$ with $v\Psi=\Phi$.
\end{theorem}

Before we come to the proof of the theorem, which can also be found in the PhD thesis of the first-named author \cite{vanluijkEntanglementNeumannAlgebraic2025}, let us briefly discuss its consequences.
Recall that every irreducible subfactor $M_A\subset M_B'$ gives rise to a bipartite system where local tomography holds, but Haag duality does not.
Since irreducible subfactors exist, see, e.g., \cite{jones_introduction_1997}, we conclude:
 If $M_A, M_B$ are commuting \emph{factors} on $\H$ then
\stackMath
        \begin{align}
            \text{local tomography}\quad \stackanchor{\Leftarrow}{\not\Rightarrow}\quad \text{Haag duality} \quad\Leftrightarrow\quad \text{Uhlmann property}. 
        \end{align}
Thus, in general, local tomography does not imply the uniqueness of purifications.

Let us illustrate how severe the failure of the Uhlmann property can be:
There exist irreducible subfactor inclusions $M_A\subset M_B'$ together with a state on $M_B$ and two purifications $\Psi, \Phi$ such that
\begin{align}
        \langle \Psi, u\Phi\rangle = 0\qquad \text{for all}\ u\in M_A.
\end{align}
We discuss a physical example below. 
Note that, as a corollary, we also see that Uhlmann's theorem fails in the strongest possible form:
The fidelity between any state on $M_B$ and itself is $1$ (one can define fidelity for general von Neumann algebras \cite{hiai_quantum_2021}), but the maximum overlap one can achieve with local unitaries of Alice between the two purifications $\Psi$ and $\Phi$ is zero.

\subsection*{Example: Surface code}
\label{sec:example}
\begin{figure}
    \centering
    \includegraphics[height=3.5cm]{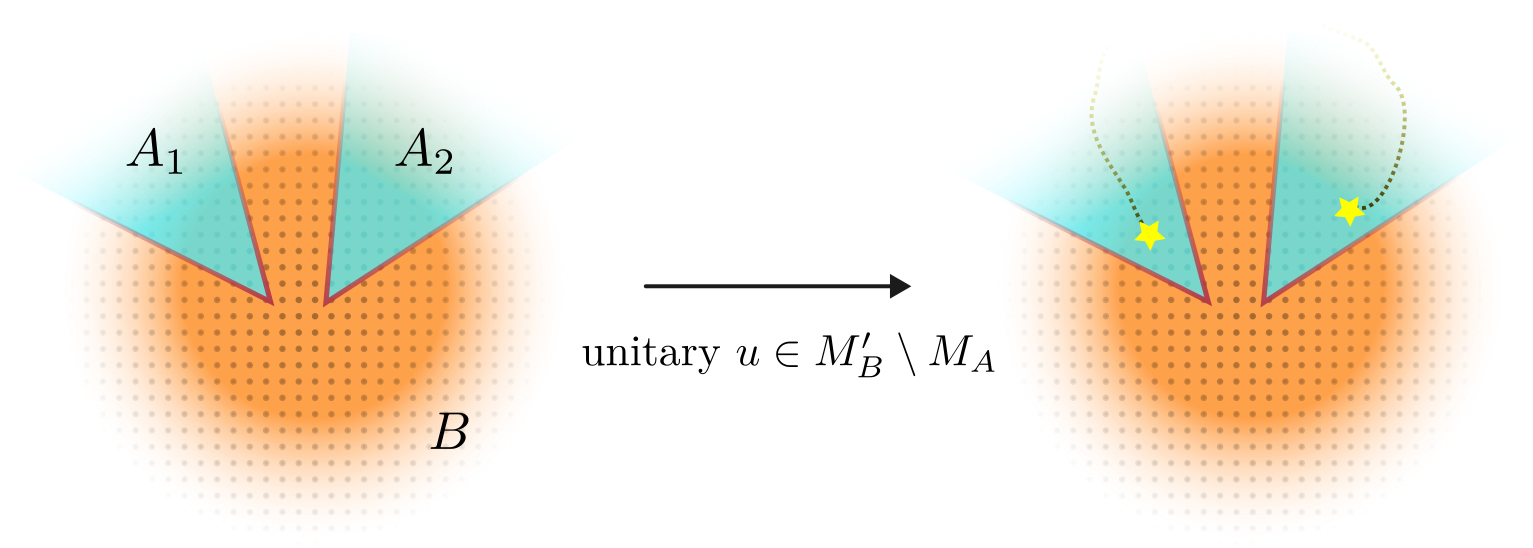}
    \caption{Surface code on the infinite lattice: A pair of anyons can be created by a unitary that is an element of the commutant of $M_B$, but not part of $M_A$, because the two regions $A_1$ and $A_2$ are not connected.}
    \label{fig:surface-code}
\end{figure}
We now illustrate the points that we made above with a physical example observed by P. Naaijkens \cite{naaijkens_kosaki-longo_2013}. 
We will restrict ourselves to physical arguments; the mathematical arguments can be found in \cite{naaijkens_kosaki-longo_2013, fiedler_jones_2017,naaijkens_subfactors_2018}. 
Consider the (unique) translation-invariant ground state of Kitaev's surface code \cite{naaijkensQuantumSpinSystems2017,naaijkensAnyonsInfiniteQuantum2012,} on an infinite square lattice $\Gamma$. We divide the lattice into three regions $A_1,A_2$ and $B = (A_1\cup A_2)'$, where the prime denotes the complement of a set of lattice sites, as shown in \cref{fig:surface-code}.
Consider the Hilbert space $\H$ arising from the GNS representation of the local algebra $\mc A_\Gamma$ relative to the translation invariant ground state of the surface code. 
To every region $\Lambda$ of the lattice, we associate the von Neumann algebra $M_\Lambda$ 
generated by operators with finite support in $\Lambda$ (cf. \cref{sec:many-body}).

First, consider the bipartition $A_1$ vs $A_1'$. In this case Haag duality $M_{A_1'} = M_{A_1}'$ holds \cite{naaijkens_haag_2012,fiedler_haag_2015}. 
Hence, by \cref{thm:main} all purifications of the marginal on $A_1'$ are related by operators in $M_{A_1}$. 

Now consider instead the bipartition $A:=A_1\cup A_2$ vs $B=A'$. Instead of the ground state $\Psi\in \H$, we can also consider a state $\Phi\in\H$ where a pair of electric anyons is created on top of the ground state, one located in $A_1$ and one located in $A_2$ (the precise location within the regions is irrelevant), see also \cref{fig:surface-code}.

The state $\Phi\in\H$ is clearly orthogonal to the ground state, but it has the same marginal on $M_B$: For every operator $b$ with finite support in $B$, we can use a unitary string operator with support away from $b$ to create $\Phi$ from $\Psi$. 
Since the operators with finite support are (weakly) dense in $M_B$ we conclude 
\begin{align}
    \langle\Psi, b\Psi\rangle = \langle\Phi,b\Phi\rangle, \qquad b\in M_B.
\end{align}
By \cref{thm:main}, there exists a partial isometry $v\in M_B'$  such that $\Phi = v\Psi$.
It can be seen to arise from a limiting process, where a string operator connecting the two anyons is pushed to infinity \cite{fiedler_jones_2017,naaijkens_subfactors_2018}.

On the other hand, with unitary operators from $M_A$, one can arbitrarily move around the anyons within $A_1$ and $A_2$, respectively, but one cannot create or remove them.  In particular $v\notin M_A$ and 
\begin{align}
    \langle\Psi, u \Phi\rangle = 0,\qquad u\in M_A,
\end{align}
Since we can create three different kinds of anyons (electric, magnetic, or a combination of both), there are in fact four classes of pairwise orthogonal purifications of the ground state marginal on $M_B$.

\section{Proof of \cref{thm:main}}

\subsubsection*{Haag duality $\Rightarrow$ Uhlmann property}
Consider two states $\Psi$ and $\Phi$ that purify the same state $\omega$ on $M_A$. The subspaces $\H_\Psi = \overline{M_A \Psi}\subset \H$ and $\H_\Phi = \overline{M_A \Phi}\subset \H$ both provide GNS representations of $\omega$. 
By the uniqueness of the GNS representation up to unitary equivalence \cite[Thm.~I.9.14]{takesaki1}, there exists a unitary $u:\H_\Psi \to \H_\Phi$ that intertwines the two representations, i.e., fulfills
\begin{align}
    u a \Psi = a u\Psi = a \Phi,\quad a\in M_A. 
\end{align}
Extending $u$ to a partial isometry $v:\H\to \H$, we find that $v$ must commute with all elements of $M_A$. Hence $v\in M_A' = M_B$ by Haag duality. 
This shows the second claim.
Since the weak closure of the unitary group $\U(M_B)$ is the set of contractions in $M_B$ \cite[Ex.~II.1]{takesaki1}, for every $\eps>0$ there exists a unitary $u\in M_B$ with $\abs{\ip{\Psi}{u\Phi}}\ge 1-\eps$.
It follows that  the Uhlmann property is fulfilled.

\subsubsection*{Uhlmann property $\Rightarrow$ Haag duality}

Let $\U(M)$ denote the unitary group of a von Neumann algebra $M$. 
By the argument in the previous section, the Uhlmann property implies $\overline{\U(M_A)\Psi} = \overline{\U(M_B')\Psi}$ for all $\Psi\in \H$. 
Since a von Neumann algebra is spanned by its unitary group, we thus have $[M_A\Psi]=[M_B'\Psi]$ for all $\Psi\in \H$, where $[V]$ denotes the orthogonal projection onto the closed subspace generated by a subset $V\subset \H$.
Since a projection $p\in B(\H)$ is in a von Neumann algebra $M \subset B(\H)$ if and only if its range $p\H$ is $M'$-invariant, we have $[M_A\Psi]\in M_B$ for all $\Psi\in \H$.
Now consider some projection $p\in M_A'$ and let $\{\Omega_n\}$ be a basis of $p\H$. 
Then $p = [\cup_n M_A\Omega_n] = \vee_n [M_A\Omega_n] $, where $\vee_n p_n$ denotes the smallest projection $q \in B(\H)$ such that $q\geq p_n$ for all $n$. Since $[M_A\Omega_n] \in M_B$ by our previous observation it follows that $p\in M_B$ \cite[Cor.~3.7]{stratilaLecturesNeumannAlgebras2019}.
Hence, every projection $p\in M_A'$ is also an element of $M_B$ and therefore $M_B \subset M_A' \subset M_B$.

\section{Discussion and outlook}
We have shown that the uniqueness of purifications in a bipartite system modeled by commuting von Neumann algebras $M_A,M_B$ is equivalent to Haag duality $M_A = M_B'$. In the case of factors, the failure of both conditions is characterized by an irreducible subfactor inclusion $M_A \subset M_B'$. 
It is well known that there exists at most one normal conditional expectation $E:M_B'\to M_A$ in such a case. Associated with $E$ is an \emph{index}, which can be seen to characterize how much bigger $M_B'$ is compared to $M_A$ \cite{jones_index_1983,kosaki_extension_1986,longo_index_1989,longo_index_1990}.\footnote{If no normal conditional expectation exists, the index is defined to be $\infty$.} 
The index can also be understood as measuring how much more classical information can be encoded into $E$-invariant states when having access to $M_B'$ instead of $M_A$ \cite{fiedler_jones_2017,naaijkens_subfactors_2018}. 
Thus, if $E$ has operational meaning, then the index carries operational meaning. In upcoming work, we show that the existence of a conditional expectation that preserves a given marginal state can be given an operational meaning itself \cite{lvl_CE}.

An interesting open question is whether the uniqueness of purifications and Haag duality can also be related in a \emph{quantitative} way instead of a qualitative way. 
It is not difficult to see that situations as discussed in the example above, where purifications remain orthogonal under arbitrary operations by Alice, exist for arbitrarily large index.%
\footnote{All that is needed is a unitary orthonormal Pimsner-Popa basis \cite{pimsnerEntropyIndexSubfactors1986}. Examples can, for example, be constructed from outer actions of finite groups on type $\II_1$ factors, in which case the index measures the number of orthogonal purifications (as in the example of the surface code).} 
It is therefore not clear whether there is a general and meaningful quantitative connection that is measured by the index.

\paragraph{Acknowledgements.} We thank René Schwonnek and Reinhard F. Werner for interesting discussions. AS and LvL have been funded by a Stay Inspired Grant of the MWK Lower Saxony (Grant ID: 15-76251-2-Stay-9/22-16583/2022).

\fussy
\emergencystretch=1em
\printbibliography

\end{document}